\documentclass{PoS}

\title{A study of the excited radial vector meson $\rho$}

\ShortTitle{Excited radial vector meson $\rho$}

\author{\speaker{Milena Piotrowska}\thanks{A footnote may follow.}\\
        Institute of Physics, Jan Kochanowski University \textit{ul. Swietokrzyska 15, 25-406, Kielce, Poland.}\\
        E-mail: \email{milena.soltysiak@op.pl}}

\author{Francesco Giacosa\\
        Institute for Theoretical Physics, J. W. Goethe University,\textit{ Max-von-Laue-Str. 1, 60438 Frankfurt, Germany.}\\
        E-mail: \email{fgiacosa@ujk.edu.pl}}

\abstract{We study the strong and radiative decays of the radial
excitation of the $\rho$ meson, with quantum numbers $n$ $^{2S+1}L_{J}=2$
$^{3}S_{1},$ $I=1$, and denoted as $\rho_{E}$, by making use of an effective
model. We test different masses: $1.45$ GeV, corresponding to well known
assignment $\rho_{E}\equiv\rho(1450)$ state; $1.25$ GeV for the assignment $\rho
_{E}\equiv\rho(1250)$ (a resonance whose existence has not yet been
confirmed), and $1.35$ GeV, which lies just in the middle between them in
order to study the dependence of the results on the mass. The decay widths for
different decay processes and two branching ratios of the radially excited
meson $\rho_{E}$ were determined and compared to the experimental data
reported in the Particle Data Group. }

\FullConference{XVII International Conference on Hadron Spectroscopy and Structure - Hadron2017\\
		25-29 September, 2017\\
		University of Salamanca, Salamanca, Spain}

\begin{document}

\section{Introduction}
Quarks interact strongly via gluons and form color neutral hadrons. According
to the quark model, the bound state of a quark $(q)$ and an antiquark
$(\bar{q})$ is a conventional meson. Mesons can be further classified into
multiplets according to their quantum numbers. 

The ground-state vector meson $\rho(770)$ is very well known, but its radial
and orbital excitations are not yet fully clarified. Here, we focus on the
radial excitation of $\rho(770).\ $This state, called $\rho_{E},$ has quantum
numbers $I=1,J^{PC}=1^{--}$ ($I$ stands for the isospin, $J$ for the total
spin, $P$ for the parity, and $C$ for the charge conjugation). Usually, the
meson $\rho_{E}$ is assigned to $\rho(1450)$ \cite{isgur}. However, the mass
values for $\rho(1450)$ listed in the PDG vary in a wide range \cite{pdg}. In
addition,  there is evidence that another $\rho$ state, called $\rho(1250)$,
exists \cite{rupp, kaminski2, tornqvist, Aston, suro, kaminski, phdthesis},
but confirmation is still needed (this state is omitted from the PDG summary).
Moreover, it is not clarified if these two resonances would coincide or
represent two different $\rho$ states (and, in the latter case, which one
would be the radially excited $u\bar{d}$ state). 

In Ref. \cite{mpio} a Quantum Field Theoretical (QFT) model was used to
describe the strong and radiative decays of both radial and orbital excited
quark-antiquark vector mesons (see sec. \ref{sec.2} for details). In that
model, the radially excited meson $\rho_{E}$ was identified with $\rho
(1450)$.  In this work, we test how the results of that model change when we
change the mass of $\rho_{E}$. In practice we calculate the decays of the
excited state $\rho_{E}$ upon settings its mass to $1.45$ GeV (corresponding to the 
standard case, $\rho(1450),$ just as in Ref. \cite{mpio}), but also to $1.25$
GeV (corresponding to the not yet confirmed $\rho(1250)$), and also to $1.35$ GeV
(just in the middle to test the dependence). In this way, we aim to test if
$\rho(1250),$ instead of $\rho(1450)$,  could be the radial excitation of the
ground-state $\rho(770).$
\section{The effective Lagrangian}
\label{sec.2} We make use of an effective QFT Lagrangian in which we couple
the matrices $P,$ $V_{\mu},$ and $V_{E,\mu}$  that describe standard $\bar
{q}q$ mesonic nonets. In particular, $P$ stands for pseudoscalar mesons
$\{\pi,K,\eta(547),\eta^{\prime}(958)\}$, $V_{\mu}$ for the ground-state
vector mesons $\{\rho(770),K^{\ast}{892},\phi(1020),\omega(782)\}$, and
$V_{E,\mu}$ for the radially excited vector mesons $\{\rho_{E}\equiv
\rho(1450),K^{\ast}(1410),\phi(1680),\omega(1420)\}$. The Lagrangian reads:
\begin{equation}
\mathcal{L}=ig_{EPP}Tr\left(  [\partial^{\mu}P,V_{E,\mu}]P\right)
+g_{EVP}Tr\left(  \tilde{V}_{E}^{\mu\nu}\{V_{\mu\nu},P\}\right)  .\label{Lag}%
\end{equation}
The first term describes decays of $V_E$ into two pseudoscalar mesons ($PP$) and the
second into a pseudoscalar and a ground-state vector meson ($VP$). In our
model we have two coupling constants: $g_{EPP}=3.66\pm0.4$ and $g_{EVP}%
=18.4\pm3.8$, determined in Ref. \cite{mpio} by using experimental data from
PDG \cite{pdg}. The bracket $[.,.]$ refers to the usual commutator and
$\{.,.\}$ to the anticommutator. Finally, $V^{\mu\nu}=\partial^{\mu}V^{\nu
}-\partial^{\nu}V^{\mu}$ and $\tilde{V}_{E}^{\mu\nu}=\frac{1}{2}\epsilon
^{\mu\nu\alpha\beta}(\partial_{\alpha}V_{E,\beta}-\partial_{\beta}V_{E,\alpha
})$. This approach allows us to study the decays of resonances which
predominantly correspond to radially excited vector mesons. Here, we consider
different masses for the excited state $\rho_{E}$. For the results of the
remaining members of the nonet of radially excited vector mesons as well as
for orbitally excited vector mesons, see Ref. \cite{mpio}.\newline The
tree-level decay widths of the resonance $\rho_{E}$ can be derived from QFT by
making use of a standard calculation, and read: $\Gamma_{\rho_{E}\rightarrow
PP}=s_{EPP}\frac{|\vec{k}|^{3}}{6\pi m_{\rho}^{2}}\left(  \frac{g_{EPP}}%
{2}\lambda_{EPP}\right)  ^{2}\textnormal{,}$ $\Gamma_{\rho_{E}\rightarrow
VP}=s_{EVP}\frac{|\vec{k}|^{3}}{12\pi}\left(  \frac{g_{EVP}}{2}\lambda
_{EVP}\right)  ^{2}$\label{strong} for strong decay channels and $\Gamma
_{\rho_{E}\rightarrow\gamma P}=\frac{|\vec{k}|^{3}}{12\pi}\left(
\frac{g_{EVP}}{2}\frac{e_{0}}{g_{\rho}}\lambda_{E\gamma P}\right)  ^{2}$
\label{rgammap} for radiative decay channel, where $\vec{k}$ is the
three-momentum of one decay product, $s_{EPP}$ and
$s_{EVP}$ are the symmetry factors, $\lambda_{EPP}$, $\lambda_{EVP}$ and
$\lambda_{E\gamma P}$ refer to amplitude coefficients (see Ref \cite{mpio}),
$g_{\rho}$ is a constant related to the process $\rho(770) \rightarrow \pi \pi$, and $e_{0}$ is the proton electric charge.
\section{Decays of $\rho_{E}$}
\label{sec.3} In this section we present the results for the decay of the excited
radial vector meson $\rho_{E}$. We examine how the decay widths depend on the
mass of the decaying resonance $\rho_{E}$ and then we compare the theoretical
values to the experimental data. In table 1 we report the results for three
types of decays (into $PP$, $VP$ and $\gamma P$) by testing the assignments
$\rho_{E}~\equiv~\{\rho(1250),$ $\rho(1350),$ $\rho(1450)\}$. Moreover, (+)
means that decay channel was observed in experiments and (-) means that it was
not. The decay $\rho_{E}\rightarrow\omega\pi$ shows that, if we use the
presently known experimental value listed under $\rho(1450)$ in the PDG, the
agreement of theory with data is spoiled.     \begin{table}[h]
\caption{Values of decay widths for strong and radiative decays of the excited
radial vector meson $\rho_{E}$ with different masses.}%
\centering
\par
\renewcommand{\arraystretch}{1.25}
\begin{tabular}
[c]{c|c|c|c|c}\hline
& \multicolumn{4}{c}{Decay width [MeV]}\\\cline{2-5}%
Decay & \multicolumn{3}{c|}{Theory} & Experiment\\\cline{2-4}%
channel & $\rho(1250)$ & $\rho(1350)$ & $\rho(1450)$ & \\\hline
$\rho_{E} \rightarrow\bar{K}K$ & $3.2 \pm0.7$ & $4.8 \pm1.0$ & $6.6 \pm1.4$ &
$< 6.7 \pm1.0$\\
$\rho_{E} \rightarrow\pi\pi$ & $25.7 \pm5.6$ & $28.1 \pm6.1$ & $30.8 \pm6.7$ &
$\approx27 \pm4$\\
$\rho_{E} \rightarrow\omega\pi$ & $26.5 \pm11.0$ & $45.5 \pm18.9$ &
$74.7\pm31.0$ & $\approx84 \pm13$\\
$\rho_{E} \rightarrow K^{*}(892) K$ & $\approx0$ & $\approx0$ & $6.7 \pm2.8$ &
$+$\\
$\rho_{E} \rightarrow\rho(770) \eta$ & $\approx0$ & $0.75 \pm0.31$ & $9.3
\pm3.9$ & $<16.0 \pm2.4$\\
$\rho_{E} \rightarrow\rho(770) \eta^{\prime}$ & $\approx0$ & $\approx0$ &
$\approx0$ & $-$\\
$\rho_{E} \rightarrow\gamma\pi$ & $0.044\pm0.026$ & $0.056 \pm0.033$ & $0.072
\pm0.042$ & $-$\\
$\rho_{E} \rightarrow\gamma\eta$ & $0.12\pm0.07$ & $0.17 \pm0.10$ & $0.23
\pm0.14$ & $\sim0.2-1.5$\\
$\rho_{E} \rightarrow\gamma\eta^{\prime}$ & $0.013 \pm0.008$ & $0.029
\pm0.017$ & $0.056 \pm0.033$ & $-$\\\hline
\end{tabular}
\end{table}

The PDG provides also informations on experimental data about ratios of decay
widths for various decay channels. The first quantity that we check is the
$\pi\pi$ to $\omega\pi$ ratio. In the PDG we find:
\begin{equation}
\left.  \frac{\Gamma_{\rho(1450)\rightarrow\pi\pi}}{\Gamma_{\rho
(1450)\rightarrow\omega\pi}}\right\vert _{\exp}\sim0.32\hspace{0.5cm}%
\textnormal{by CLEGG94 \cite{Clegg94}.}%
\end{equation}
The corresponding theoretical values are: $0.41\pm0.20$ for $\rho_{E}%
~\equiv\rho(1450)$ (in a good agreement with CLEGG 94); $0.97\pm0.45$ for
$\rho_{E}~\equiv\rho(1250)$; $0.62\pm0.29$ for $\rho_{E}~\equiv\rho(1350)$
(the last two values are also in qualitative agreement with the experiment).
\newline For the $KK/\omega\pi$ ratio the PDG reports:
\begin{equation}
\left.  \frac{\Gamma_{\rho(1450)\rightarrow KK}}{\Gamma_{\rho(1450)\rightarrow
\omega\pi}}\right\vert _{\exp}<0.08\hspace{0.5cm}%
\textnormal{DONNACHIE91 \cite{Donnachie91},}%
\end{equation}
which is compatible with our results for the assignment $\rho_{E}\equiv
\rho(1450)$: $0.088\pm0.043$. However, the experimental value is not
compatible for the other two masses: $0.121\pm0.057$ for $\rho_{E}~\equiv
\rho(1250)$ and $0.105\pm0.049$ for $\rho_{E}~\equiv\rho(1350)$. Both values
are too large of at least a factor 5. The results show that $\rho(1450)$ is
predominantly the radial excitation of the $\rho(770)$ meson.
\section{Conclusions}
In this work we described the strong and the radiative decays of the excited
radial vector meson $\rho_{E}$ upon using different masses. To this end a
relativistic QFT model based on flavor symmetry was employed. This model has been
previously used for both radially and orbitally excited vector mesons
\cite{mpio}. We have shown that the results for the resonance $\rho(1450)$ is in
good agreement with available experiments and with the assignment of being
(predominantly) the radial excitation of the $\rho(770)$ meson. A lighter mass,
such as a hypothetical $\rho(1250)$ meson, would be problematic (if we keep
using the decay rates presently listed under $\rho(1450)$). An interesting
future study along this direction is the evaluation of the effects of the
quantum mesonic fluctuations on the radially excited vector meson. 
\section*{Acknowledgements}

The authors thank C. Reisinger for cooperation and G. Rupp and S. Coito for discussion. The authors acknowledge support from the Polish National Science Centre (NCN) through the OPUS project no. 2015/17/B/ST2/01625.

\end{document}